\documentclass[journal=ancac3,manuscript=article,layout=onecolumn]{achemso}

\usepackage{amssymb}
\usepackage{amsmath}

\usepackage{microtype}

\usepackage[tx]{sfmath}

\title{Nonlocal Response of Metallic Nanospheres Probed by Light, Electrons, and Atoms}

\keywords{Nonlocal response, nanoplasmonics, EELS, extinction, LDOS, spontaneous emission, multipole plasmons.}

\author{Thomas~Christensen}
\affiliation{Department of Photonics Engineering, Technical University of Denmark, DK-2800 Kgs. Lyngby, Denmark}
\alsoaffiliation{Center for Nanostructured Graphene, Technical University of Denmark, DK-2800 Kgs. Lyngby, Denmark}
\author{Wei~Yan}
\affiliation{Department of Photonics Engineering, Technical University of Denmark, DK-2800 Kgs. Lyngby, Denmark}
\alsoaffiliation{Center for Nanostructured Graphene, Technical University of Denmark, DK-2800 Kgs. Lyngby, Denmark}
\author{S\o ren~Raza}
\affiliation{Department of Photonics Engineering, Technical University of Denmark, DK-2800 Kgs. Lyngby, Denmark}
\alsoaffiliation{Center for Electron Nanoscopy, Technical University of Denmark, DK-2800 Kgs. Lyngby, Denmark}
\author{Antti-Pekka~Jauho}
\affiliation{Department of Micro- and Nanotechnology, Technical University of Denmark, DK-2800 Kgs. Lyngby, Denmark}
\alsoaffiliation{Center for Nanostructured Graphene, Technical University of Denmark, DK-2800 Kgs. Lyngby, Denmark}
\author{N.~Asger~Mortensen}
\affiliation{Department of Photonics Engineering, Technical University of Denmark, DK-2800 Kgs. Lyngby, Denmark}
\alsoaffiliation{Center for Nanostructured Graphene, Technical University of Denmark, DK-2800 Kgs. Lyngby, Denmark}
\author{Martijn~Wubs}
\email{mwubs@fotonik.dtu.dk}
\affiliation{Department of Photonics Engineering, Technical University of Denmark, DK-2800 Kgs. Lyngby, Denmark}
\alsoaffiliation{Center for Nanostructured Graphene, Technical University of Denmark, DK-2800 Kgs. Lyngby, Denmark}

\begin{document}

\begin{abstract}
Inspired by recent measurements on individual metallic nanospheres that can not be explained with traditional classical electrodynamics, we theoretically investigate the effects of nonlocal response by metallic nanospheres in three distinct settings: atomic spontaneous emission, electron energy loss spectroscopy, and light scattering. These constitute two near-field and one far-field measurements, with {zero-,} {one-,} and two-dimensional excitation sources, respectively.
We search for the clearest signatures of hydrodynamic pressure waves in nanospheres. We employ a linearized hydrodynamic model and Mie--Lorenz theory is applied for each case.
Nonlocal response shows its mark in all three configurations, but for the two near-field measurements we predict especially pronounced nonlocal effects that are not exhibited in far-field measurements. Associated with every multipole order is not only a single blueshifted surface plasmon, but also an infinite series of bulk plasmons that has no counterpart in a local-response approximation. We show that these increasingly blueshifted multipole plasmons become spectrally more prominent at shorter probe-to-surface separations and for decreasing nanosphere radii. For selected metals we predict hydrodynamic multipolar plasmons to be measurable on single nanospheres.

\end{abstract}

\textbf{Keywords:} Nonlocal response, nanoplasmonics, EELS, extinction, LDOS, spontaneous emission, multipole plasmons.
\vskip 1em \hrule\vskip 1em
%
%
A plethora of effects arises in structured metals due to collective excitations of conduction electrons and their interaction with the electromagnetic field. This constitutes plasmonics, a research field with mature roots\cite{Ritchie:1957,Pines:1952} that is continuing to develop strongly.\cite{Stockman:2011} Notably, applications for plasmonics are found in the biochemistry and biomedical fields, \textit{e.g.}, in surface-enhanced Raman spectroscopy (SERS),\cite{Campion:1998} biosensing\cite{VazquezMena:2011} and biomedical imaging,\cite{Khlebtsov:2010} drug delivery,\cite{Kyrsting:2010} and phototherapy of cancer-cells.\cite{Lal:2008} Purely photonic applications are also emerging, \textit{e.g.}, in plasmonic waveguiding,\cite{Bozhevolnyi:2006b} optical nanoantennas\cite{Novotny:2011,Muskens:2007}, and photovoltaics.\cite{Wu:2011}

Recent years' advances in fabrication, synthesis, and characterization techniques have allowed well-controlled experimental investigations of plasmonics even at the nanoscale. Yet in this growing field of nanoplasmonics,\cite{Stockman:2011,Stockman:2011_OptExpress} the commonly employed theory for light-matter interaction is still traditional classical electrodynamics, where the response of the material constituents to light is described collectively in terms of local, bulk-material response-functions. Indeed, this approach usually remains very accurate, even for sub-wavelength phenomena.

Interestingly, recent measurements on individual few-nanometer plasmonic particles have shown phenomena that are clearly beyond classical electrodynamics. Electron energy-loss spectroscopy (EELS) of Ag spheres resting on dielectric substrates showed surface plasmon resonance blueshifts up to $0.5\,\mathrm{eV}$ as compared to classical theory\cite{Scholl:2012,Raza:2013_Nanophotonics}. Earlier similar measurements were performed on ensembles of nanoparticles.\cite{vomFelde:1988} Classical electrodynamics was also shown to fail in experiments involving (sub-)nanometer-sized gaps between dimers,\cite{Kern:2012,Savage:2012,Scholl:2013a} or between nanoparticles and a substrate.\cite{Ciraci:2012}

To explain these features arising beyond the validity of classical electrodynamics, various physical mechanisms are invoked. Firstly, classical electrodynamics assumes a step-function profile of the free-electron density at a metal-dielectric interface. The finite quantum mechanical spill-out\cite{Lang:1970a} of the electron density redshifts the surface plasmon resonance,\cite{Liebsch:1993a,Teperik:2013} may give rise to nonresonant field enhancement,\cite{Ozturk:2011} and may enable charge transfer between non-touching plasmonic dimers.\cite{Esteban:2012,Savage:2012,Scholl:2013a}
Secondly, a stronger confinement of the free electrons gives rise to blueshifts. In cluster physics, it is single-particle excitations that are blueshifted due to quantum confinement,\cite{DeHeer:1993} while confinement in nanoplasmonics blueshifts collective resonances and gives rise to Friedel oscillations in the electron density.\cite{Keller:1993,Townsend:2011a}
A third, semi-classical physical mechanism beyond classical electrodynamics is nonlocal response, discussed in more detail below, which becomes important when reducing the particle size or gap size of a dimer down to the range of the nonlocality\cite{Ginzburg:2013} ($\xi_{\textsc{nl}}$, denoting the spatial extent of significant nonlocal interaction, to be introduced shortly), and blueshifts surface plasmon resonance frequencies.

Large experimental blueshifts of the localized surface plasmon (LSP) dipole resonance seem to indicate that several physical mechanisms add up.\cite{Raza:2013_Nanophotonics,Raza:2013_OE}
Certainly, in experiments all these physical mechanisms beyond traditional classical electrodynamics are at work simultaneously, thus emphasizing the importance of microscopic theories\cite{Stella:2013} (\textit{e.g.}, density-functional theory, DFT) or effective models\cite{CarminaMonreal:2013a} that incorporate multiple mechanisms. Yet at the same time it is important to ascertain the relative strength and compatibility of the various mechanisms. Indeed, it is paramount to know - and to measure - the unique characteristics of each mechanism, that is to say, find their individual ``smoking guns'', in order to appreciate the dominant physical mechanisms under different nonstandard circumstances. We foresee an increasing number of such decisive experiments on individual nanoparticles in the near future.

The boundary between cluster physics and nanoplasmonics is an interesting one. Metal clusters require a quantum description of interacting electron states, often studied with DFT. In contrast, nanoplasmonics could be defined to start for nanoparticle sizes that allow an effective quantum description in terms of non-interacting plasmons\cite{Townsend:2011a}. A current interesting issue is where to place the origin of the observed blueshift of the surface plasmon resonance of individual nanospheres: is it primarily due to quantum confinement of single-particle states,\cite{vomFelde:1988,Scholl:2012} or due to confinement of collective modes?\cite{Keller:1993,Townsend:2011a,Raza:2013_Nanophotonics,CarminaMonreal:2013a} In this article we assume the latter and identify new observable consequences. We focus on nanoparticles that are considered large enough ($2R \geq 3$ nm) that so-called core plasmons, although collective in nature, can be neglected according to DFT calculations.\cite{Townsend:2011a}

Nonlocal response is a semi-classical effect which emerges in nanoplasmonics at few-nanometer length scales. The general nonlocal relation between the displacement and electric fields, $\mathbf{D}(\mathbf{r},\omega) = \varepsilon_0 \int\varepsilon(\mathbf{r},\mathbf{r}';\omega) \mathbf{E}(\mathbf{r}',\omega) \,\mathrm{d}\mathbf{r}'$ becomes simpler and more familiar in the local-response approximation (LRA), \textit{i.e.}, $\varepsilon(\mathbf{r},\mathbf{r}';\omega) \simeq \varepsilon_{\textsc{lra}}(\mathbf{r},\omega)\delta(\mathbf{r}-\mathbf{r}')$. In many cases this approximation provides an excellent effective description due to the short-range nature of the nonlocal interaction. However, the LRA is not justifiable when the nonlocal interaction length, $\xi_{\textsc{nl}}$, becomes comparable with characteristic feature sizes of structural or optical kind.\cite{Ginzburg:2013}

Here we consider inclusion of the classically neglected Fermi--Dirac pressure of the electron gas. Its associated pressure waves give rise to a nonlocal optical response. The simplest way to study the effects of Fermi pressure in nanoplasmonics is by assuming a hydrodynamic model,\cite{Ruppin:1973a,David:2012,Raza:2013_Nanophotonics,Fuchs:1987,Baltz:1995,TranThoai:1986,TranThoai:1988,Mortensen:2013} which neglects the aforementioned spill-out and confinement effects on the static electron density. In hydrodynamics the nonlocal interaction length becomes $\xi_{\textsc{nl}} = v_{\textsc{f}}/\omega$, with $v_{\textsc{f}} = \hbar\sqrt[3]{3\pi^2n_0}/m$ denoting the Fermi velocity, defined through the effective mass $m$ and free-electron density $n_0$. This corresponds to $\xi_{\textsc{nl}}$-values in the range $2\,\textrm{-}\,5\,\textrm{\AA}$ for typical plasmonic metals at optical frequencies, see Table S1 in the Supporting Information (SI). We will focus on the linearized hydrodynamic model here, but would like to emphasize that the full hydrodynamic model involves both nonlocality and nonlinearity, predicting nonlinear effects such as second-harmonic generation at the surface of metal nanoparticles for larger field-strengths\cite{Sipe:1980,Ginzburg:2012_PRB,Ciraci:2012_PRB}.

The strongest evidence of hydrodynamic behavior in metals originate from experiments on thin metal films, where resonances due to standing waves of confined bulk plasmons have been identified, in silver by Lindau and Nilsson\cite{Lindau:1971}, in potassium by Anderegg \textit{et al.}\cite{Anderegg:1971}, in magnesium by Chen\cite{Chen:1976}, and very recently by {\" O}zer \textit{et al.}\cite{Oezer:2011} Rather surprisingly, {\" O}zer \textit{et al.}\cite{Oezer:2011} could measure confined bulk plasmon resonances (\textit{i.e.}, standing Fermi pressure waves) even for ultrathin magnesium films of only three atomic monolayers, and found qualitative agreement with theory even when neglecting electronic spill-out. For nanospheres on the other hand, the observations of blueshifted dipole-resonances of localized surface plasmons (LSPs) in individual nanospheres\cite{Scholl:2012,Raza:2013_Nanophotonics,Raza:2013_OE} and of broad resonance-features above the plasma frequency in ensembles,\cite{Duthler:1971} tentatively suggested as associated with confined bulk plasmons,\cite{Ruppin:1973a} are perhaps less conclusive evidence of hydrodynamic behavior. This may in part be due to a line of reasoning which addresses just a single resonance, namely the dipole.

Our aim in this article is then to examine theoretically which phenomena constitute the clearest evidence of hydrodynamic pressure waves in plasmonic nanospheres, and how best to observe them.
Powerful measurement techniques include scattering measurements, as realized \textit{e.g.}, in the infrared regime by Fourier transform infrared spectroscopy (FTIR), scanning near-field optical microscopy (SNOM)\cite{Greffet:1997}, EELS\cite{Abajo_PhysRevMod2010,Egerton:2009}, and fluorescence microscopy techniques, utilizing decay enhancement of emitters near plasmonic resonances\cite{Schmelzeisen:2010,Willets:2013}.
In this theoretical article, we systematically explore three prominent measurement techniques, each with different excitation sources, namely the extinction cross-section, the EELS probability, and the electric local density of states (LDOS). The excitation sources are, respectively, a linearly polarized plane wave, a traveling electron with kinetic energy in the $\mathrm{keV}$-range, and an electric dipole emitter, corresponding to a two-, one-, and zero-dimensional source. The three measurement principles represent both far- and near-field types, and we show their spectra to be qualitatively different.

We investigate not only the strongest (dipolar) LSP resonance of nanospheres, but also higher-order multipole LSPs, as well as bulk plasmons, for all three measurements considered.
We show that hydrodynamic response leads to a significant spectral separation of the sphere's multipole plasmons at small radii, allowing them to extend above the LRA asymptotic limit at $\omega_{\mathrm{p}}/\sqrt{2}$. Resonance-features above this limit have already been observed in polydisperse ensembles of nanospheres, and previously been interpreted instead in terms of single-particle confinement.\cite{vomFelde:1988} We find significant qualitative disparity between properties measurable in the far-field, \textit{i.e.}, \textit{via} extinction, and in the near-field, \textit{i.e.}, \textit{via} EELS or LDOS.
Our findings result in concrete suggestions to experimentally observe hydrodynamic nonlocal phenomena in the near-field, by identifying the multipolar plasmon resonances of individual nanospheres of selected metals.

%
%
\section{Results and Discussion}
\paragraph{Theoretical framework.}
In a linearized hydrodynamic description, the current density $\mathbf{J}(\mathbf{r},\omega)$ and the electric field $\mathbf{E}(\mathbf{r},\omega)$ are interrelated by the nonlocal relation\cite{Boardman:1982a,Raza:2011}:
\begin{subequations}\label{eqs:governing}
\begin{equation}\label{eq:hydromain}
\frac{\beta_{\textsc{f}}^2}{\omega(\omega+i\eta)}\nabla[\nabla\cdot \mathbf{J}(\mathbf{r},\omega)] + \mathbf{J}(\mathbf{r},\omega) = \sigma(\omega)\mathbf{E}(\mathbf{r},\omega),
\end{equation}
where $\sigma(\omega) = i\varepsilon_0\omega_\mathrm{p}^2/(\omega+i\eta)$ is the usual Drude conductivity of a free-electron gas with plasma frequency $\omega_{\mathrm{p}}$, including a phenomenological loss-rate $\eta$, and $\beta_{\textsc{f}}^2 = (3/5)v_{\textsc{f}}^2$ is the hydrodynamical velocity of plasma pressure waves in the metal. The hydrodynamic model can be classified as `semi-classical' because Eq.~(\ref{eq:hydromain}) relates the classical fields $\mathbf{J}$ and $\mathbf{E}$ \textit{via} the parameter $\beta_{\textsc{f}}\propto v_{\textsc{f}}$ which is proportional to $\hbar$. Hydrodynamic response appears as a lowest order spatially nonlocal correction to the local Ohm's law, with a strength proportional to $\xi_{\textsc{nl}}^{-2}k^2$ in momentum $k$-space.

In addition to Eq.~\eqref{eq:hydromain}, the electric field must satisfy the Maxwell wave equation
\begin{equation}\label{eq:vectorwave}
\nabla\times\nabla\times\mathbf{E}(\mathbf{r},\omega) - k_0^2\varepsilon_{\infty}(\omega)\mathbf{E}(\mathbf{r},\omega) = i\omega\mu_0 \mathbf{J}(\mathbf{r},\omega),
\end{equation}
\end{subequations}
with $k_0 = \omega/c$ denoting the usual free-space wavenumber, and $\varepsilon_{\infty}(\omega)$ the dielectric response of the bound charges, \textit{i.e.}, the response not due to the free-electron plasma.
The sum of the bound- and free-electron response gives the transverse response of the metal $\varepsilon_{\textsc{m}}(\omega) = \varepsilon_{\infty}(\omega)+\sigma(\omega)/i\varepsilon_0\omega$, familiar from the LRA.
For calculations involving a measured transverse metal response $\varepsilon_{\textsc{m}}(\omega)$, the bound response $\varepsilon_{\infty}(\omega)$ is determined by fixing $\omega_{\mathrm{p}} = \sqrt{n_0 e^2/\varepsilon_0 m}$, \textit{i.e.}, through the free-electron density $n_0$ and effective mass $m$, thus determining the free response $\sigma(\omega)$ and allowing $\varepsilon_{\infty}(\omega)$ to be determined by subtraction.\cite{David:2012}.

The practical solution of Eqs.~\eqref{eqs:governing} in structures with curvilinear symmetries can be aided significantly by expansion in the so-called vector wave functions. Concretely, a monochromatic electromagnetic field in a region of uniform dielectric function, can be expanded in the basis composed of the solenoidal, $\mathbf{M}_{\nu}(\boldsymbol{\mathrm{r}})$ and $\mathbf{N}_{\nu}(\boldsymbol{\mathrm{r}})$, and irrotational, $\mathbf{L}_{\nu}(\boldsymbol{\mathrm{r}})$, vector wave functions: \cite{Chew:book,Stratton}
\begin{equation}
\boldsymbol{\mathrm{E}}(\boldsymbol{\mathrm{r}}) =
\sum_{\nu} a_{\nu}\mathbf{M}_{\nu}(\boldsymbol{\mathrm{r}})
+b_{\nu}\mathbf{N}_{\nu}(\boldsymbol{\mathrm{r}})
+c_{\nu}\mathbf{L}_{\nu}(\boldsymbol{\mathrm{r}}),
\end{equation}
where $\nu$ denotes a composite expansion index with $a_{\nu}$, $b_{\nu}$, and $c_{\nu}$ being associated expansion coefficients. The functions $\mathbf{M}_{\nu}(\boldsymbol{\mathrm{r}})$ and $\mathbf{N}_{\nu}(\boldsymbol{\mathrm{r}})$ describe the TE and TM parts, respectively, of the electric field, and describe the propagation of transverse, or divergence-free, modes.\cite{Stratton} The functions $\mathbf{L}_{\nu}(\boldsymbol{\mathrm{r}})$ are irrotational, and as such are irrelevant in media described by the LRA. However, their inclusion is indispensable for the treatment of plasmonic nanoparticles by hydrodynamic response, in order to account for the inclusion of longitudinal modes.

\begin{figure}\centering
\includegraphics[scale=1.18]{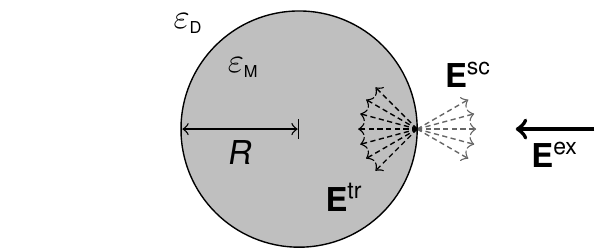}
\caption{Sketch of an exciting wave $\boldsymbol{\mathrm{E}}^{\mathrm{ex}}$ interacting with a metallic sphere embedded in a dielectric background, giving rise to scattered and transmitted fields $\boldsymbol{\mathrm{E}}^{\mathrm{sc}}$ and $\boldsymbol{\mathrm{E}}^{\mathrm{tr}}$, respectively.}\label{fig:schematic}
\end{figure}

Next, we consider the case of an arbitrary external exciting field $\boldsymbol{\mathrm{E}}^{\mathrm{ex}}$ that originates in an outer dielectric region and scatters upon a spherical metallic particle of radius $R$ that is centered at the origin. This induces scattered fields $\boldsymbol{\mathrm{E}}^{\mathrm{sc}}$ outside the particle and transmitted fields $\boldsymbol{\mathrm{E}}^{\mathrm{tr}}$ inside, see Figure~\ref{fig:schematic}. For spherical nanoparticles, the choice of multipolar vector wave functions separates the composite expansion index $\nu$ into the angular-momentum quantum numbers $l$ and $m$, for details see the Methods section.

Outside the nanosphere ($r>R$), the fields $\boldsymbol{\mathrm{E}}^{\mathrm{ex}}$ and $\boldsymbol{\mathrm{E}}^{\mathrm{sc}}$ can be expanded solely in terms of the in- and outgoing transverse multipoles $\{\mathbf{M}_{lm}^{\rm ex},\mathbf{N}_{lm}^{\rm ex}\}$ and $\{\mathbf{M}_{lm}^{\rm sc},\mathbf{N}_{lm}^{\rm sc}\}$, respectively, since the dielectric region does not support longitudinal waves. The corresponding expansion coefficients are $\{a_{lm}^{\mathrm{ex}},b_{lm}^{\mathrm{ex}}\}$ and $\{a_{lm}^{\mathrm{sc}},b_{lm}^{\mathrm{sc}}\}$. The transmitted field $\boldsymbol{\mathrm{E}}^{\mathrm{tr}}$ inside the nanosphere ($r<R$) requires besides ingoing transverse multipoles, $\{\mathbf{M}_{lm}^{\rm tr},\mathbf{N}_{lm}^{\rm tr}\}$, also ingoing longitudinal modes $\mathbf{L}_{lm}^{\mathrm{tr}}$, which correspondingly necessitates three sets of expansion coefficients $\{a_{lm}^{\mathrm{tr}},b_{lm}^{\mathrm{tr}},c_{lm}^{\mathrm{tr}}\}$.

The fields inside and outside the nanosphere are related by boundary conditions (BCs), see the Methods section. This translates into linear relations between the expansion coefficients of the exciting and scattered fields\cite{Stratton,Quinten_2011}
\begin{equation}\label{eq:exc_vs_sca}
a_{lm}^{\mathrm{sc}} = t_{l'}^{\textsc{te}} a_{l'm'}^{\mathrm{ex}}\delta_{ll'}\delta_{mm'}, \qquad
b_{lm}^{\mathrm{sc}} = t_{l'}^{\textsc{tm}} b_{l'm'}^{\mathrm{ex}}\delta_{ll'}\delta_{mm'},
\end{equation}
where $\delta_{jk}$ is the Kronecker delta.
The proportionality constants $t_l^{\textsc{te}}$ and $t_l^{\textsc{tm}}$ are known as the Mie--Lorenz coefficients\cite{Mie:1908}. For nanospheres with nonlocal response they are given by\cite{Ruppin:1973a,David:2012}
\begin{subequations}\label{eqs:Miecoefs}
\begin{align}
t_l^{\textsc{te}} &= \frac
{-j_l(x_{\textsc{m}})[x_{\textsc{d}}j_l(x_{\textsc{d}})]'+j_l(x_{\textsc{d}})[x_{\textsc{m}}j_l(x_{\textsc{m}})]'}
{j_l(x_{\textsc{m}})[x_{\textsc{d}}h_l^{\scriptscriptstyle (1)}(x_{\textsc{d}})]'-h_l^{\scriptscriptstyle (1)}(x_{\textsc{d}})[x_{\textsc{m}}j_l(x_{\textsc{m}})]'}, \label{eq:Miecoefs_TE}\\
t_l^{\textsc{tm}} &= \frac
{-\varepsilon_{\textsc{m}} j_l(x_{\textsc{m}})[x_{\textsc{d}}j_l(x_{\textsc{d}})]'+\varepsilon_{\textsc{d}}j_l(x_{\textsc{d}})\big\lbrace [x_{\textsc{m}} j_l(x_{\textsc{m}})]'+\Delta_l\big\rbrace}
{\varepsilon_{\textsc{m}} j_l(x_{\textsc{m}})[x_{\textsc{d}}h_l^{\scriptscriptstyle (1)}(x_{\textsc{d}})]'-\varepsilon_{\textsc{d}}h_l^{\scriptscriptstyle (1)}(x_{\textsc{d}})\big\lbrace [x_{\textsc{m}} j_l(x_{\textsc{m}})]'+\Delta_l\big\rbrace },\label{eq:Miecoefs_TM}
\end{align}
where $x_{\textsc{d}} = k_{\textsc{d}}R$ and $x_{\textsc{m}} = k_{\textsc{m}}R$ are dimensionless parameters in terms of the dielectric and transverse metal wavenumbers (see Methods), and the radius $R$ of the nanosphere. The primes denote the derivatives with respect to $x_{\textsc{d,m}}$.
As for the usual Mie--Lorenz coefficients in the LRA, these hydrodynamic Mie--Lorenz coefficients are independent of the multipole label $m$, due to the spherical geometry of the scatterer.
Spatial nonlocality influences the Mie--Lorenz coefficients through the hydrodynamic term\cite{Ruppin:1973a,David:2012}
\begin{equation}\label{eq:Mie_HydroCorrec}
\Delta_l = l(l+1) j_l(x_{\textsc{m}})\frac{\varepsilon_{\textsc{m}}-\varepsilon_{\infty}}{\varepsilon_{\infty}}\frac{j_l(x_{\textsc{nl}})}{x_{\textsc{nl}}j_l'(x_{\textsc{nl}})},
\end{equation}
\end{subequations}
with $x_{\textsc{nl}} = k_{\textsc{nl}}R$ introducing the longitudinal metal wavenumber (see Methods). As expected, the correction $\Delta_l$ vanishes in the LRA limit, since $|x_{\textsc{nl}}|\rightarrow \infty$ as $\beta_{\textsc{f}}\rightarrow 0$.
Note that only the scattering of TM waves is affected by the inclusion of spatial nonlocality. There are no contributions to the magnetic field from the longitudinal multipoles $\mathbf{L}_{lm}^{\rm tr}$, \textit{cf.}\ the Maxwell--Faraday equation, thus leaving the TE waves, sometimes called the magnetic waves, unaffected.

The significance of the Mie--Lorenz coefficients is that they specify the scattering laws outside the sphere, \textit{i.e.}, they determine the outcome of external measurements. In particular, a general linear measurement $\mathcal{O}$ on a nanosphere can be expressed as a linear combination of them.
As discussed in more detail below, all three measurements that we consider can be expressed in the general form
\begin{equation}\label{eq:O_measurement_general}
\mathcal{O} = \sum_{lm} \mathcal{O}_{lm}^{\textsc{te}}\mathrm{Re}(t_l^{\textsc{te}}) + \mathcal{O}_{lm}^{\textsc{tm}}\mathrm{Re}(t_l^{\textsc{tm}}),
\end{equation}
where the coefficients $\mathcal{O}_{lm}^{\textsc{te},\textsc{tm}}$ contain all information regarding the measurement, \textit{e.g.}, type and position, while $t_l^{\textsc{te},\textsc{tm}}$ contain all information regarding the scattering geometry, \textit{e.g.}, dielectric composition and size. Crucially, the inclusion of hydrodynamic nonlocality modifies only the Mie--Lorenz coefficients $t_l^{\textsc{tm}}$ -- but not the measurement coefficients $\mathcal{O}_{lm}^{\textsc{te},\textsc{tm}}$.

For this reason we can first focus on the Mie--Lorenz coefficients and look for the local and nonlocal plasmonic resonances that in principle affect all measurements. After that, we will identify the measurements in which these resonances make a prominent appearance and where the impact of hydrodynamic dispersion is especially strong.

\begin{figure}
\hspace*{-0.3cm}\includegraphics[]{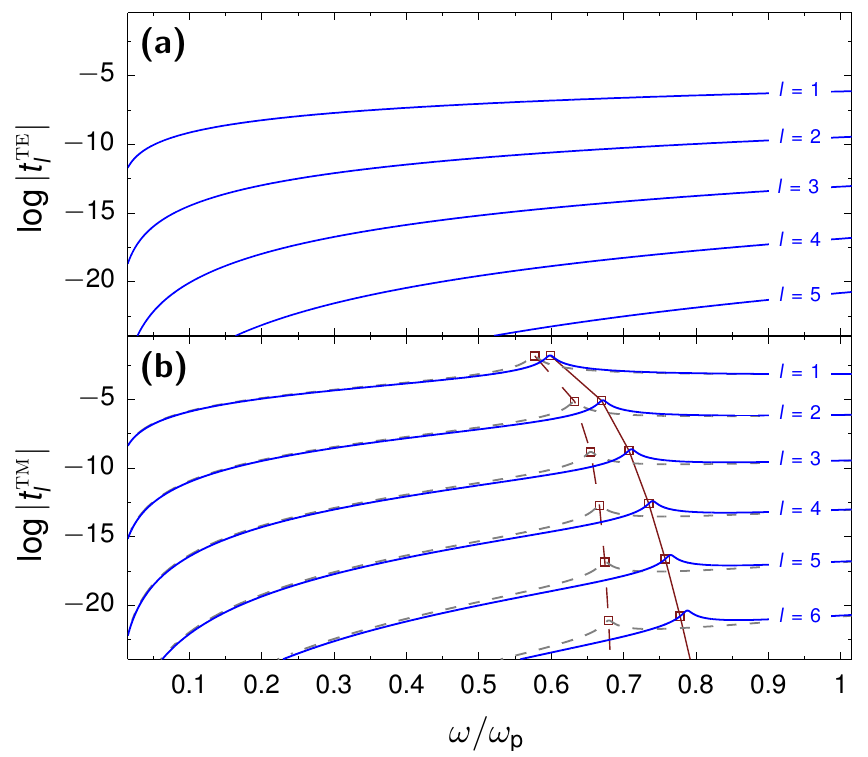}
\caption{Absolute value of the Mie--Lorenz coefficients $t_l^{\textsc{te}}$ and $t_l^{\textsc{tm}}$ in (a) and (b), respectively, on a logarithmic scale, as a function of frequency, for the first few values of $l$. Considered is a $R=2.5\, \mathrm{nm}$ sphere with Drude-metal parameters $\omega_{\mathrm{p}} = 10\, \mathrm{eV}$, $\eta = 0.1\,\mathrm{eV}$ and $\varepsilon_{\infty} = 1$ embedded in vacuum, $\varepsilon_{\textsc{d}} = 1$. For comparison, the LRA TM Mie--Lorenz coefficients are illustrated in gray dashed lines. Approximate resonance predictions for LRA and hydrodynamics, as predicted by Eqs.~\eqref{eq:localresonances} and~\eqref{eq:approxmultipoleres}, are given in dashed and full red lines, respectively.}\label{fig:MieCoefs}
\end{figure}

\paragraph{Multipole plasmon resonances.}
%
Figure~\ref{fig:MieCoefs} depicts the frequency dependence of the first few Mie--Lorenz coefficients $t_l^{\textsc{te},\textsc{tm}}$ of a free-electron $R = 2.5\,\mathrm{nm}$ nanosphere. Clearly, large-$l$ multipoles in general scatter significantly weaker than small-$l$ multipoles (notice the log scale). In addition, the $t_l^{\textsc{tm}}$ coefficients exhibit a series of resonances, corresponding to poles of the coefficient, associated with excitation of LSPs of dipole, quadrupole, hexapole (and so on) character, for $l=1,2,3,\ldots$, respectively. By contrast, the $t_l^{\textsc{te}}$ coefficients exhibit no such resonances. Moreover they are several orders of magnitude smaller than their equal-momenta TM correspondents. As a result, the TM-interaction dominates the response of plasmonic nanospheres. It is this dominant TM-interaction which is modified by nonlocal response.

\emph{Surface plasmon resonance conditions.}\quad{}
%
A trademark of hydrodynamic response is its blueshift of resonances as compared to local response. Figure~\ref{fig:MieCoefs} illustrates that for nanospheres these blueshifts show up in the TM Mie--Lorenz coefficients, and are increasingly shifted for larger $l$.\cite{Yan:2013_GSIM} We study this quantitatively and find the multipole plasmon resonances of order $l$ from the pole of the $t_l^{\textsc{tm}}$ coefficient. The nonretarded limit can be applied to the small spheres under consideration, leading to the plasmon condition\cite{BoardmanParanjape:1977}
\begin{equation}\label{eq:multipolecondition}
l\varepsilon_{\textsc{m}} + (l+1)(1+\delta_l)\varepsilon_{\textsc{d}} = 0,
\end{equation}
where $\delta_l=\Delta_l/[j_l(x_{\textsc{m}})(l+1)]$ accounts for the hydrodynamic correction, see SI for additional details (a similar multipole plasmon condition was derived in Ref.~\citenum{BoardmanParanjape:1977} for metallic spheres in vacuum, but with a missing factor of $i/x_{\textsc{nl}}$ in their equivalent definition of $\delta_l$). Evidently, nonlocality can be interpreted as modifying the dielectric surrounding, by introducing an effective $l$-dependent dielectric constant $\varepsilon_{l,\textsc{d}}^{\mathrm{eff}} = (1+\delta_l)\varepsilon_{\textsc{d}}$. Since $\delta_l$ itself is a function of frequency and angular momentum, Eq.~\eqref{eq:multipolecondition} defines plasmon resonances only implicitly. Nevertheless, their spectral location can be determined by approximation while retaining the essential physics, as we shall show below.

In the LRA limit $\delta_l \rightarrow 0$ and upon neglecting dispersion of the bound response and damping, \textit{i.e.}, taking $\varepsilon_{\textsc{m}}(\omega) = \varepsilon_{\infty}-\omega_{\mathrm{p}}^2/\omega^2$, the well-known local electrostatic plasmon resonances are immediately recovered from Eq.~\eqref{eq:multipolecondition} as
\begin{equation}\label{eq:localresonances}
\omega_l^{\textsc{l}} = \frac{\omega_{\mathrm{p}}}{\sqrt{\varepsilon_{\infty}+\tfrac{l+1}{l}\varepsilon_{\textsc{d}}}},
\end{equation}
Thus, in local theory, for $l=1$ we find the well-known (dipolar) LSP resonance $\omega_l^{\textsc{l}} = \omega_{\mathrm{p}}/\sqrt{\varepsilon_{\infty}+2\varepsilon_{\textsc{d}}}$, which reduces to $\omega_{\mathrm{p}}/\sqrt{3}$ for a free Drude-metal sphere in vacuum. The high-order multipole plasmons tend asymptotically from below towards the local planar-interface surface plasmon $\omega_{\mathrm{p}}/\sqrt{\varepsilon_{\infty}+\varepsilon_{\textsc{d}}}$ for $l\rightarrow \infty$, reducing to $\omega_{\mathrm{p}}/\sqrt{2}$ for a free Drude-metal sphere in vacuum. The $l$-dependence of $\omega_l^{\textsc{l}}$ as described by Eq.~\eqref{eq:localresonances} is depicted by the red-dashed line in Figure~\ref{fig:MieCoefs}, clearly showing the asymptotic behavior for large $l$.

Turning now from local to nonlocal response, let us assume that $\delta_l$ in Eq.~\eqref{eq:multipolecondition} is a small perturbation, which is valid for small $l$ and for $R\gg\beta_{\textsc{f}}/\omega_{\mathrm{p}}$. We circumvent the implicitness of the resonance condition by making a pole approximation, replacing the dispersive function $\delta_l(\omega)$ by its value $\delta_l^{\textsc{l}} = \delta_l(\omega_l^{\textsc{l}})$ in the local resonance frequency $\omega_l^{\textsc{l}}$, the latter given by Eq.~\eqref{eq:localresonances}. The hydrodynamically corrected resonances $\omega_l^{\textsc{nl}}$ then occur at approximately\cite{Yan:2013_GSIM}
\begin{equation}\label{eq:approxmultipoleres}
\omega_l^{\textsc{nl}}
\simeq \frac{\omega_{\mathrm{p}}} {\sqrt{ \varepsilon_{\infty} + \tfrac{l+1}{l}(1+\delta_l^{\textsc{L}})\varepsilon_{\textsc{d}}}}
%
%
\simeq \omega_l^{\textsc{l}} + \frac{\beta_{\textsc{f}}}{R}\sqrt{\frac{l(l+1)\varepsilon_{\textsc{d}}}{4\varepsilon_{\infty}}},
\end{equation}
where, at the last step, in addition to a Taylor expansion of the square-root term, we have utilized the large imaginary $x_{\textsc{nl}}$ limit of the hydrodynamic correction, $\delta_l \simeq l\tfrac{\varepsilon_{\textsc{m}}-\varepsilon_{\infty}}{\varepsilon_{\infty}}\tfrac{i}{x_{\textsc{nl}}}$, which is applicable at frequencies below the screened plasma frequency $\omega_{\mathrm{p}}^{\scriptscriptstyle \infty} \equiv \omega_{\mathrm{p}}/\varepsilon_{\infty}$. These approximate nonlocal surface plasmon resonance frequencies are illustrated by the solid red line in Figure~\ref{fig:MieCoefs}. The approximation captures the exact nonlocal blueshift well but is less accurate for larger $l$, as expected. By implication of these nonlocal blueshifts, excitations appear between the LRA $l= \infty$ mode (the planar surface plasmon) and the volume plasmon at $\omega_{\mathrm{p}}$, classically a resonance-free frequency interval.\cite{vomFelde:1988}

\emph{Bulk plasmon resonance condition.}\quad{}
%
Besides blueshifting the multipolar LSP resonances that already exist in the LRA, hydrodynamical theory also predicts the appearance of additional resonances due to confined bulk plasmons for which no LRA counterparts exist.\cite{Ruppin:1973a,Raza:2011} More microscopic theories have also predicted the emergence of such bulk plasmons.\cite{Townsend:2011a,Stella:2013} These bulk plasmons emerge due to the presence of propagating, longitudinal pressure waves above the plasma frequency. In hydrodynamics, the confined bulk plasmons are then easily interpreted as the standing-wave resonances of longitudinal waves. Table~\ref{tab:multipoles} depicts isosurfaces of the induced charge density for LSPs and bulk plasmons for comparison.
\begin{table}
\includegraphics[]{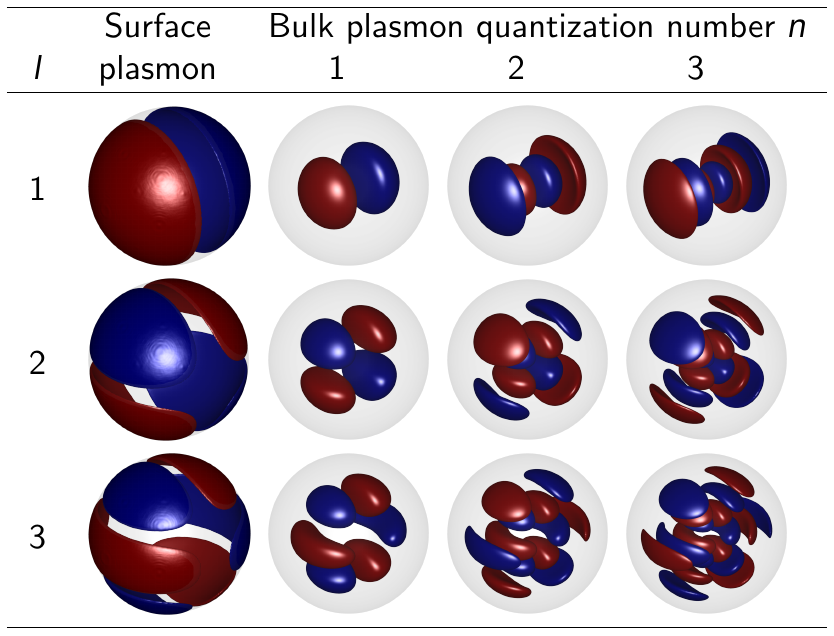}
\caption{Charge densities of multipole surface and bulk plasmons. Isosurfaces are drawn for the real part of the charge density, calculated in a hydrodynamic treatment, at isovalues equal to plus/minus (red/blue) twice the mean of the absolute value of the charge density in the sphere. The nanosphere outline is indicated in shaded gray.\label{tab:multipoles}}
\end{table}

An approximation for these bulk resonances can be found by neglecting the coupling of the pressure waves to light, \textit{i.e.}, by searching for standing wave solutions of $\mathbf{L}_{lm}^{\mathrm{tr}}$, thus neglecting the transverse components. For nanospheres, this gives radially quantized confined bulk plasmons resonating at the frequencies $\omega_{ln}^{\mathrm{bulk}}$ (see SI for details):
\begin{equation}\label{eq:bulkplasmonapproximation}
\omega_{ln}^{\mathrm{bulk}}(\omega_{ln}^{\mathrm{bulk}} + i \eta) = \frac{\omega_{\mathrm{p}}^{2}}{\varepsilon_{\infty}} + w_{ln}^{2}\bigg(\frac{\beta_{\textsc{f}}}{R}\bigg)^{2},
\end{equation}
where $w_{ln}$ is the $n^{th}$ positive root of $j_{l}'(w)$, the derivative of the $l^{\mathrm{th}}$-order spherical Bessel function (see Refs.~\citenum{Ruppin:1973a} and~\citenum{Gildenburg:2011} for lengthier, more accurate approximations). Modes associated with the first root at $n=0$ are in fact not resonant, but are artifacts of the approximation that arise due to having neglected the transverse field-components.
Regardless, for every multipole order $l$ there is an infinite number of confined bulk plasmons associated with $n=1,2,\ldots$.

\begin{figure}
\hspace*{-0.3cm}\includegraphics[]{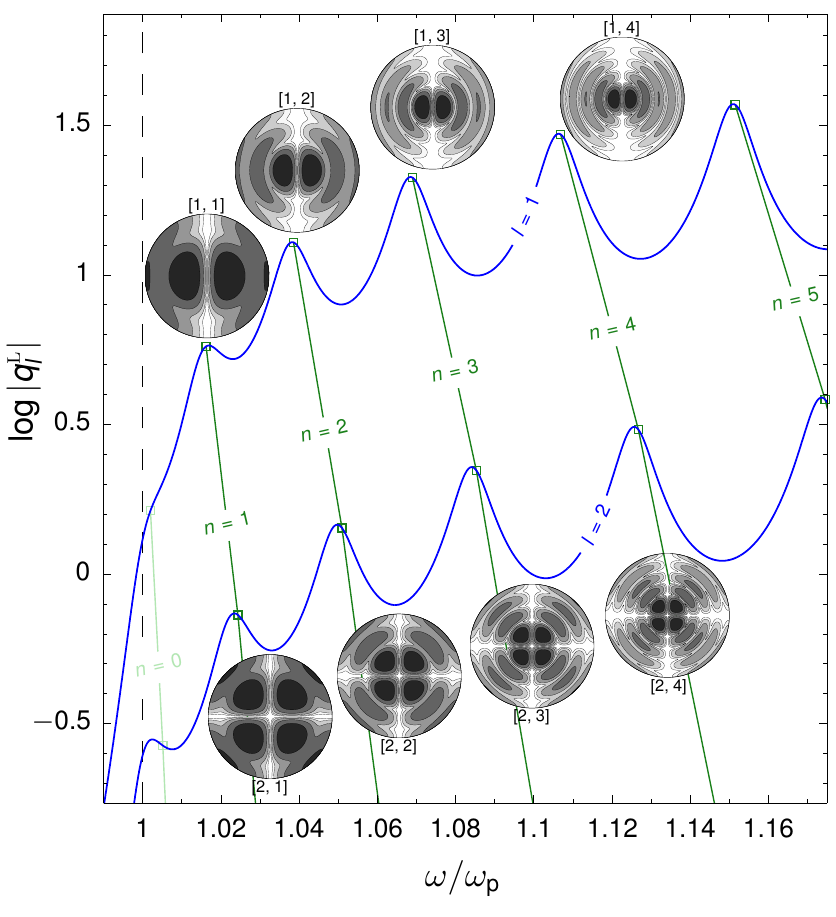} 
\caption{Absolute value of the Mie--Lorenz transmission coefficients $q_{l}^{\textsc{l}}$ on a logarithmic scale, as a function of frequency. The coefficients give the coupling amplitude between transmitted longitudinal multipoles, and incident TM multipoles. Setup-parameters are identical to those in Figure~\ref{fig:MieCoefs}. Shown are the dipolar, $q_{1}^{\textsc{l}}$ and the quadrupolar, $q_{2}^{\textsc{l}}$, coefficients in blue. Both exhibit peaks above $\omega_{\mathrm{p}}$, corresponding to a series of confined bulk plasmons labeled by $n=0,1,2,\ldots$. Green curves show approximate resonance positions, see Eq.~\eqref{eq:bulkplasmonapproximation}. The absence of an $n=0$ resonance is apparent. Insets depict logarithmic scale contour plots, with contours separated by factors of 2, of the absolute value of the induced charge density of the bulk resonances, with $[l,n]$ indices labeled, in the $xz$-plane.}\label{fig:MieCoefs_BulkFocus_Insets}
\end{figure}

As for the LSP resonances, we first illustrate the signature of these bulk plasmons in the Mie--Lorenz coefficients, before considering the experiments in which their presence is most pronounced.
In Figure~\ref{fig:MieCoefs_BulkFocus_Insets} we depict the frequency dependence of the first few Mie--Lorenz transmission coefficients $q_{l}^{\textsc{l}}$ near and above $\omega_{\mathrm{p}}$. These coefficients give the transmission amplitude to a longitudinal mode due to excitation by an incident TM mode, and are defined analogously to the scattering coefficients $t_{l}^{\textsc{te},\textsc{tm}}$ of Eq.~\eqref{eq:exc_vs_sca} through $c_{lm}^{\mathrm{tr}} = q_{l'}^{\textsc{l}} b_{l'm'}^{\mathrm{ex}}\delta_{ll'}\delta_{mm'}$, see SI for their explicit form.
%
%
The first dipolar and quadrupolar bulk plasmon resonances of a nanosphere clearly show up as Lorentzian resonances, and the bulk plasmon approximation Eq.~\eqref{eq:bulkplasmonapproximation} is quite accurate. The resonant charge distributions in the insets illustrate the radial quantization of the confined bulk plasmons.
To the best of our knowledge, only the dipole ($l=1$) confined bulk plasmons have been considered previously, \textit{e.g.}, in relation with extinction-features above the plasma frequency in nanospheres.\cite{Ruppin:1973a,Raza:2011} In our investigation of EELS and LDOS below, we consider additionally if these higher-$l$ bulk plasmons may influence the spectral response in the near field. First, however, we discuss the properties of higher-order LSP multipoles.

\begin{figure}
\hspace*{-0.3cm}\includegraphics[]{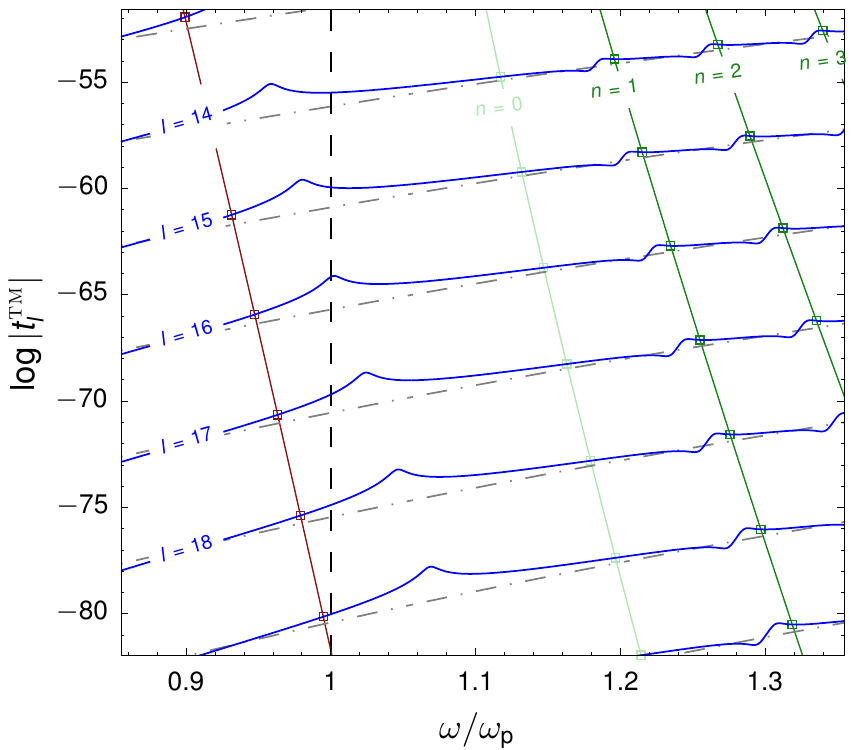} 
\caption{Absolute value of the TM Mie--Lorenz coefficients, $t_l^{\textsc{tm}}$, on a logarithmic scale, as a function of frequency for high-angular momenta. Setup-parameters are identical to those in Figure~\ref{fig:MieCoefs}. Hydrodynamic results are illustrated in blue solid lines, while LRA results are illustrated by gray dashed lines for comparison. The transition across the plasma frequency is marked by the black dashed line. The red line depicts the approximate LSP resonance of Eq.~\eqref{eq:approxmultipoleres}; the green lines show Eq.~\eqref{eq:bulkplasmonapproximation} and approximate the first few confined bulk plasmon resonances. The bulk plasmons show up as Fano-like resonances in $|t_l^{\textsc{tm}}|$.\cite{Tribelsky:2012}}\label{fig:MieCoefs_bulkvsvolume}
\end{figure}

\emph{Large-$l$ plasmonic resonances.}\quad{}
We have seen in Figure~\ref{fig:MieCoefs} that multipolar hydrodynamic LSP modes blueshift away from the classical limit, the LRA planar surface plasmon at $\omega_{\rm p}/\sqrt{2}$.
What is more, Figure~\ref{fig:MieCoefs_bulkvsvolume}
illustrates that high-multipole nonlocal LSP resonances can even appear above the plasma frequency $\omega_{\rm p}$. There is no indication that the plasma frequency would mark a qualitative transition. This is despite the change from predominantly imaginary metal wavenumbers ($k_{\textsc{m}}$ and $k_{\textsc{nl}}$) for frequencies $\omega<\omega_{\mathrm{p}}^{\infty}$, to predominantly real metal wavenumbers for $\omega > \omega_{\mathrm{p}}^{\infty}$. In particular, the transition from predominantly imaginary to real wavenumbers does not carry with it a transition from predominantly bound surface modes to volume-like modes as assumed in the past.\cite{Baltz:1995}
[Such a transition does not emerge since $|x_{\textsc{nl}}|$ remains comparative with $\sqrt{l+1}$, which, \textit{cf.}\ Eq.~\eqref{eqs:SphericalVectorWaves} and the small-argument asymptotic form $j_l(x)\simeq x^l/(2l+1)!!$ valid for $|x|\ll\sqrt{l+1}$, implies that $|j_l(x)|\sim |j_l(ix)|$ for $|x|<|x_{\textsc{nl}}|$, whereby the charge density is left qualitatively unchanged and surface-bound.]
Hydrodynamic surface plasmons above the plasma frequency have also been found theoretically for a planar metal-dielectric interface, for a thin metal slab, and for planar metamaterials.\cite{Wei_PRB2012,Raza:2013_PhysRevB}

It is fruitful to pursue further the analogy between the LSPs of our nanospheres and of planar structures. The analogy is well-known for local response, but the hydrodynamic version holds a surprise. The large-$l$ LSP resonances below and above the plasma frequency can both be characterized by wave propagation along the surface of the nanosphere.
The $l^{\mathrm{th}}$ surface mode accommodates exactly $l$ oscillation periods along the periphery of the sphere. One can therefore ascribe an effective surface wavelength $\lambda^{\mathrm{s}}_l = 2\pi R/l$ and an effective surface wavenumber $k^{\mathrm{s}}_l = l/R$ to the $l^{\mathrm{th}}$ mode. For larger $l$, the effective wavelength becomes shorter and the modes perceive the curving surface of the sphere as increasingly flat. For that reason the dispersion would mimic that of a planar metal-dielectric interface for large $l$.

\begin{figure}
\centering
\hspace*{-.3cm}\includegraphics[]{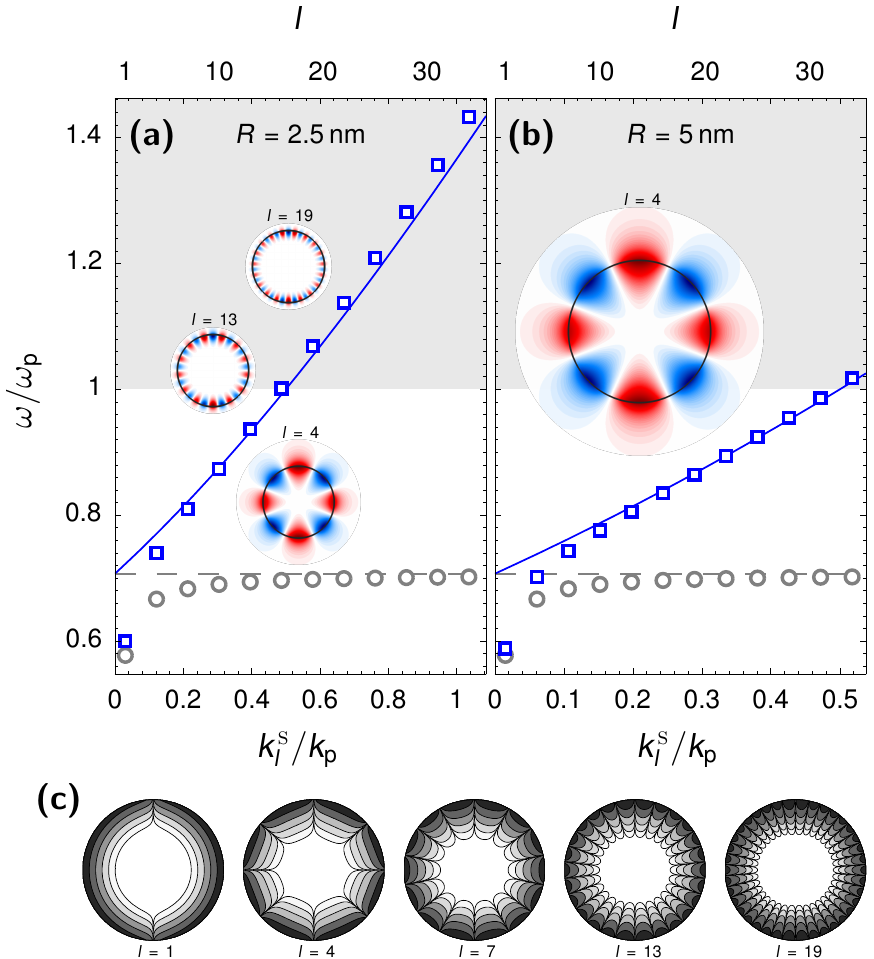} 
\caption{Dispersion of the nonretarded surface plasmon resonances of nanospheres. Material-parameters as in Figure~\ref{fig:MieCoefs} but with $\eta = 0$. Wavenumbers are normalized to the plasma wavenumber $k_{\mathrm{p}} = \omega_{\mathrm{p}}/\beta_{\textsc{f}}$. The hydrodynamic model is shown in blue and the LRA in gray. The $l=1,4,7,\ldots,34$ multipole LSP resonances are indicated by squares and circles; nonretarded dispersion relations\cite{Raza:2013_PhysRevB} for a planar interface are shown as solid lines. Insets in panels (a) and (b) show the real parts of the electric field of selected LSP modes in the $xz$-plane along $\theta$-polarization (on separate color scales). Panel (c) depicts contour plots of the absolute value of the hydrodynamic charge density of selected LSP modes in the same nanosphere and in the same plane (contours separated by factors of 10 with separate, logarithmic color scales).
}\label{fig:pseudodispersion}
\end{figure}

To test this prediction from the analogy, we compute the exact plasmon resonances from Eq.~\eqref{eq:multipolecondition} and show them in a pseudo-dispersion plot in Figure~\ref{fig:pseudodispersion}.
%
For local response, Figure~\ref{fig:pseudodispersion} indeed shows the well-known result that for larger $l$ the dispersion of the nanosphere LSPs approaches more and more that of a flat interface.
For nonlocal response, also shown in Figure~\ref{fig:pseudodispersion}, we first note that the LSP dispersion indeed does not show a transition at the plasma frequency, as we already guessed from Figure~\ref{fig:MieCoefs_bulkvsvolume}. Secondly, there is satisfactory agreement of the hydrodynamic dispersion of LSPs for a nanosphere and for the flat interface, so the analogy is also meaningful for hydrodynamic response. However, and this is the surprising third point, unlike for local response, the agreement does {\em not} converge towards a complete agreement as $l$ increases: a discrepancy develops for large $l$. The discrepancy is larger in Figure~\ref{fig:pseudodispersion}(a) for $R=2.5\,{\rm nm}$ spheres than for the twice larger spheres in Figure~\ref{fig:pseudodispersion}(b).

This can be explained by noting that in the LRA all the induced free charge resides only \emph{on} the surface of the sphere, whereas it is distributed \emph{close to} this surface in the hydrodynamic description. The latter is illustrated in Figure~\ref{fig:pseudodispersion}(c). Note the surficial standing-wave quantization of the LSPs in Figure~\ref{fig:pseudodispersion}, and also the absence of radial quantization, being associated only with the bulk plasmons as shown in Figure~\ref{fig:MieCoefs_BulkFocus_Insets}. For large $l$, the neighboring hydrodynamic charge patterns in Figure~\ref{fig:pseudodispersion} get squeezed into each other due to the finite curvature, producing the discrepancy with the planar interface. An alternative explanation of the discrepancy as due to interaction across antipodal surface points can be ruled out, since the insets of Figures~\ref{fig:pseudodispersion}(a,b) show that the electric fields corresponding to high-$l$ modes are well localized near the surface of the nanosphere, even those above the plasma frequency (in contrast to predictions of Ref.~\citenum{Baltz:1995}), so that fields on opposite angular regions of the sphere are spatially well separated. This agrees with recent findings for hydrodynamic LSP modes in a planar thin metal slab, which do not show finite-size effects either for sufficiently large wavevectors. Rather, since the slab has no curvature, the large-$k$ dispersion of its LSP modes does indeed agree with that of the single interface.\cite{Wei_PRB2012}

\paragraph{Extinction, EELS, and LDOS.}
%
Having discussed the characteristics of the multipole plasmons, and in particular the modifications due to hydrodynamic response, we will now consider three distinct measurements, each with a different sensitivity to the various surface and bulk plasmons:
\begin{enumerate}
\item \textbf{Light scattering.} This measurement gives the extinction cross-section $\sigma_{\mathrm{ext}}(\omega)$, yielding the ratio of power dissipated due to scattering and absorption of a plane-wave relative to incident intensity.
\item \textbf{Electron energy-loss spectroscopy. } EELS gives information on the electron loss function $\Gamma(\omega)$, that expresses the probability that a relativistic electron will lose an energy $\hbar\omega$ due to interaction with the particle. We consider electrons traveling with velocity $v\hat{\mathbf{z}}$ and impact parameter $\mathbf{b}$ in the $xy$-plane outside the sphere ($|\mathbf{b}|=b>R$).
\item \textbf{Atomic spontaneous emission.} A dipole orientation-averaged measurement of local spontaneous emission rates relates linearly to the electric local optical density of states (or LDOS) $\rho^{\textsc{e}}(\omega)$. We consider emitter positions $\mathbf{b}$ outside the nanosphere ($b > R$).
\end{enumerate}
These three measurements constitute examples of illumination of the sphere by plane-, cylinder-, and spherical-like waves. Extinction is measured in the archetypical far-field scattering setup, while the EELS probability and LDOS can be measured locally in the near-field.
Sub-nanometer control of the probe-surface separation is routinely achieved in EELS\cite{Abajo_PhysRevMod2010} and also demonstrated in fluorescence measurements\cite{Anger:2006,Dulkeith:2005}, permitting experimental investigation of the various calculated spectra that we will show below.

Let us briefly discuss the computation of these measurements in the multipole basis.
The arbitrary exciting field can be decomposed into the multipole basis, \textit{i.e.}, the coefficients $\{a_{lm}^{\mathrm{ex}},b_{lm}^{\mathrm{ex}}\}$ can be determined. The scattered field is then obtained through the Mie--Lorenz coefficients using Eq.~\eqref{eq:exc_vs_sca}. A general linear measurement $\mathcal{O}$ may involve components of the scattered field at a single location, as for the LDOS, or a continuous weighting of different spatial components of the field, as for the extinction cross-section or the EELS probability. In any case, the measurements can be expressed through a weighted $lm$-summation of the scattering amplitudes $t_l^{\textsc{te}} a_{lm}^{\mathrm{ex}}$ and $t_l^{\textsc{tm}} b_{lm}^{\mathrm{ex}}$. As stated above, for the extinction cross-section\cite{BohrenHuffman:1983}, EELS probability\cite{Abajo:1999,Abajo_PhysRevMod2010}, and LDOS\cite{Kerker:1980,Ruppin:1982,Chew:1987,Dung:2001pra,Vos:2009}, the measurements $\mathcal{O}$ can all be expressed in terms of the Mie--Lorenz coefficients in the general form of Eq.~\eqref{eq:O_measurement_general}.
For the specific forms that Eq.~\eqref{eq:O_measurement_general} takes for each of the three measurements, we refer the reader to Eqs.~(S3), (S6) and~(S9) of the SI.

In the following we normalize the extinction cross-section to the geometric cross-section, $\pi R^2$, yielding the extinction efficiency $Q_{\mathrm{ext}}(\omega) \equiv \sigma_{\mathrm{ext}}(\omega)/\pi R^2$, and similarly normalize the LDOS to the free-space LDOS $\rho^{\textsc{e}}_0(\omega)$, yielding the LDOS enhancement $[\rho^{\textsc{e}}/\rho^{\textsc{e}}_0](\omega)$.

\begin{figure*}
\hspace*{-.5cm}
\makebox[\textwidth]{
\includegraphics{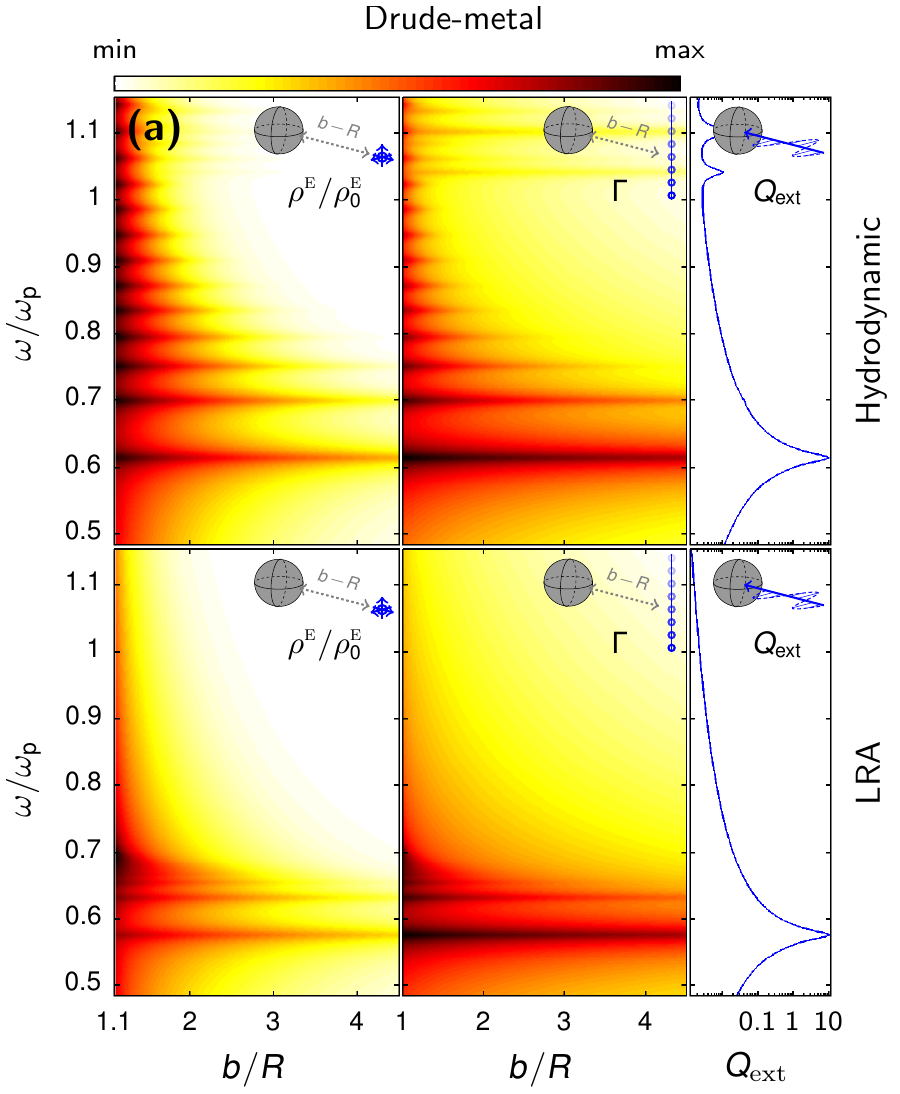}
\includegraphics{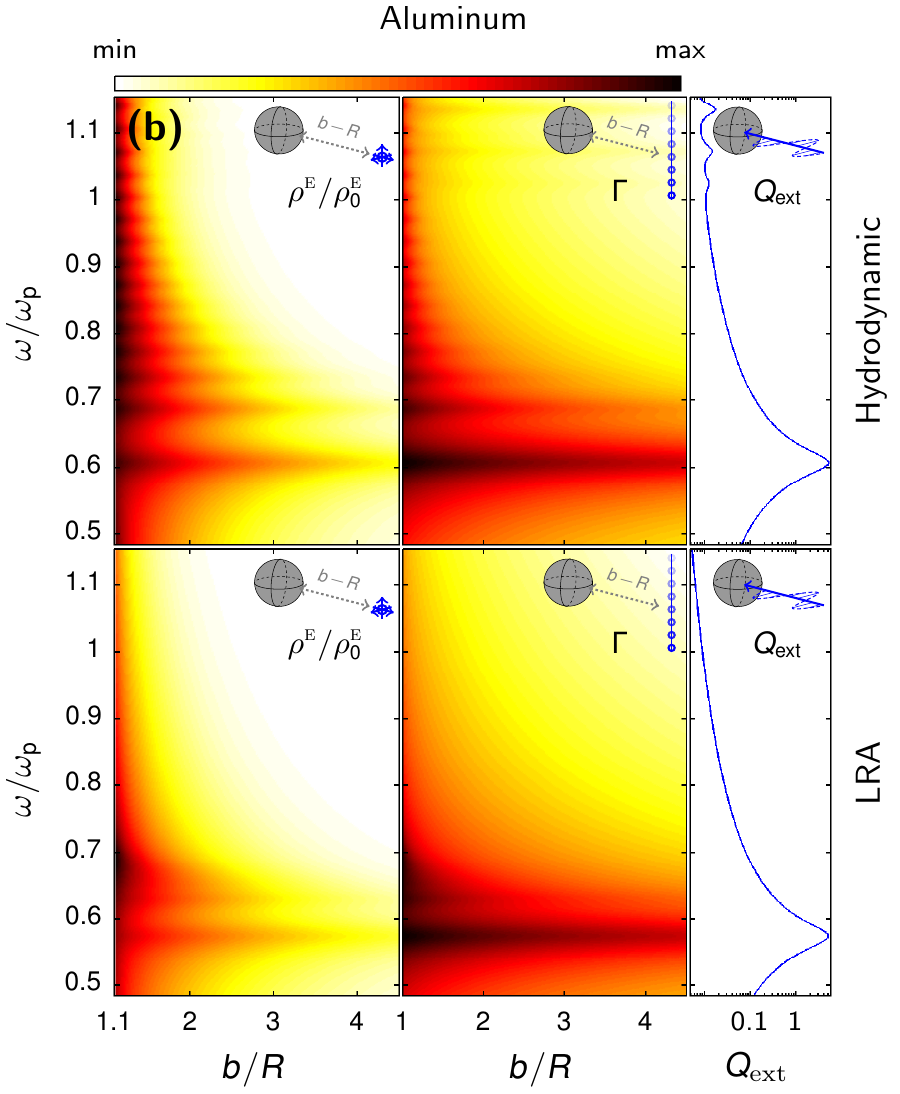}}
\caption{Normalized LDOS, $\rho^{\textsc{e}}/\rho^{\textsc{e}}_0$, EELS probability, $\Gamma$, and extinction efficiency, $Q_{\mathrm{ext}}$, in left, center, and right panels, respectively. LDOS and EELS calculations are illustrated on independent logarithmic color scales. An $R = 1.5\, \mathrm{nm}$ sphere in vacuum is considered. Electron energy in EELS calculations is $E_e = 200\, \mathrm{keV}$. (a) Drude-metal with $\omega_\mathrm{p} = 10\, \mathrm{eV}$, $\eta = 0.1\, \mathrm{eV}$, and $\varepsilon_{\infty} = 1$. (b) Aluminum with bound response included from measured data from Ref.~\citenum{Rakic:1995} \textit{via} $\varepsilon_\infty(\omega)$, with $\omega_\mathrm{p} = 14.94\, \mathrm{eV}$ and $\eta = 0.075\, \mathrm{eV}$. }\label{fig:EELSLDOSExt}
\end{figure*}

\begin{figure*}
\hspace*{-1cm}
\makebox[\textwidth]{
\includegraphics{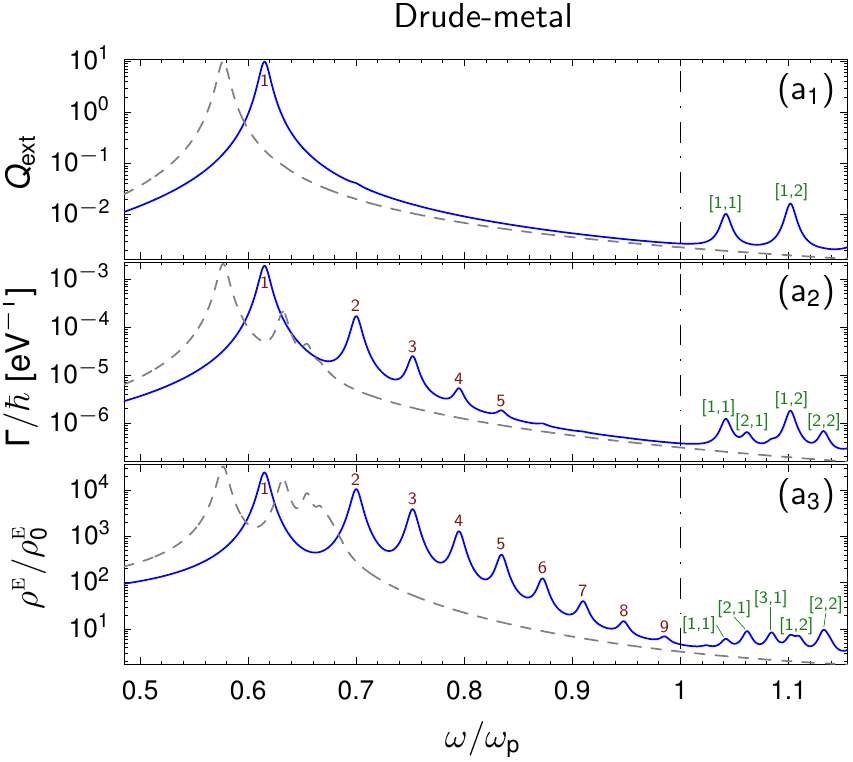}
\includegraphics{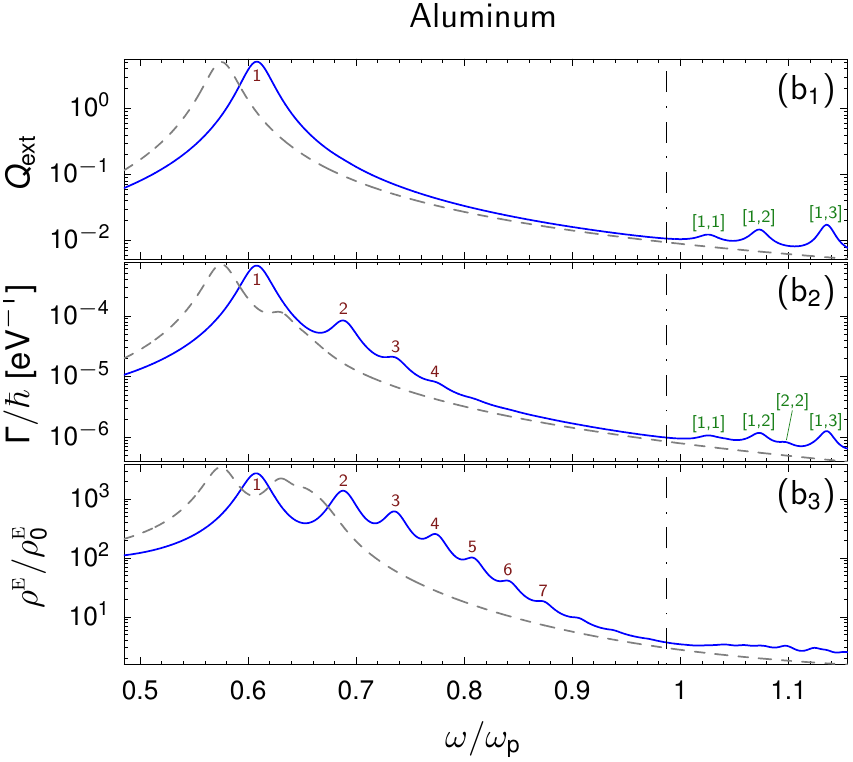}}
\caption{Extinction efficiency, $Q_{\mathrm{ext}}$, EELS probability, $\Gamma$, and normalized LDOS, $\rho^{\textsc{e}}/\rho^{\textsc{e}}_0$. Hydrodynamics in full blue and LRA in dashed gray. The screened plasma frequency is indicated in dashed-dotted black. When distinguishable, the LSP multipole order $l$ is noted in red, while bulk plasmon $[l,n]$-orders are noted in green. Parameters in (a) are as in Figure~\ref{fig:EELSLDOSExt}(a); in (b) as in Figure~\ref{fig:EELSLDOSExt}(b). The EELS probability and LDOS are computed for $b/R=2$ in all three cases.}\label{fig:ExtEELSLDOS_1Dplot}
\end{figure*}

\emph{Near-field versus far-field.}\quad{}
%
Figure~\ref{fig:EELSLDOSExt}(a)
depicts the probe-to-surface separation dependence of the LDOS and EELS spectra in a Drude-metal nanosphere of $R=1.5\,\mathrm{nm}$, and for comparison also depicts the extinction resonances. Hydrodynamic and LRA calculations are shown to be distinctly different. Most conspicuous in Figure~\ref{fig:EELSLDOSExt}(a) is perhaps that many new resonances appear in the nonlocal EELS and LDOS spectra, many more than in extinction, and that drastic changes occur when we vary $b/R$ from the contact scenario $b/R=1$ to $b/R=4.5$. When fixing $b/R=2$, we obtain the spectra of Figure~\ref{fig:ExtEELSLDOS_1Dplot}(a).
Below we discuss both figures in more detail, but before that, Figure~\ref{fig:EELSLDOSExt}(a) already makes clear that only a rudimentary understanding of EELS measurements can be obtained by comparing them with calculated extinction or absorption spectra. Such comparisons have nevertheless been quite common.\cite{Scholl:2012,Raza:2013_Nanophotonics}

Let us interpret Figures~\ref{fig:EELSLDOSExt}(a) and~\ref{fig:ExtEELSLDOS_1Dplot}(a) in more detail by first discussing the region below the plasma frequency, where both in local and nonlocal response, the extinction efficiency exhibits just the single dipolar ($l=1$) surface plasmon resonance. Higher-order multipole plasmons do not contribute since the sphere size is much smaller than the wavelength of the incident plane wave.\cite{Ruppin:1973a}

In stark contrast to these known extinction spectra, several additional multipole LSP resonances are observable in the EELS and LDOS spectra, and better so for smaller probe-to-surface separations.
Notice that higher-order LSP modes do exist in the LRA, as we have seen in the analysis of the Mie--Lorenz coefficients, but these additional LSP resonances converge towards the $l=\infty$ limit at $\omega_{\mathrm{p}}/\sqrt{2}$ and rapidly become indistinguishable due to losses.
By contrast, the higher-order LSP resonances are much more clearly visible in the hydrodynamic spectra because of the $l$-dependent nonlocal blueshift of Eq.~\eqref{eq:approxmultipoleres}, which pushes the multipole resonances in the EELS and LDOS spectra beyond the LRA $l=\infty$ limit and moreover separates them despite the loss-induced broadening.\cite{TranThoai:1986,TranThoai:1988}

Observation of a multipolar resonance above the $l=\infty$ limit was reported by vom~Felde \textit{et al.}\cite{vomFelde:1988} in EELS measurements on ensembles of potassium clusters of radius $1-2\,\mathrm{nm}$ embedded in magnesium oxide. Vom~Felde \textit{et al.} attributed this blueshift into the classically quiet region to quantum size effects. Here we show that there is a good alternative explanation, namely collective hydrodynamic multipolar LSP resonances. Thus the ongoing discussion how to interpret the blueshift of the main dipolar LSP resonance as seen in EELS\cite{Scholl:2012,Raza:2013_Nanophotonics,CarminaMonreal:2013a} can now be extended to higher-order LSP resonances, observable both in EELS and LDOS measurements. This improves the outlook of obtaining conclusive evidence for hydrodynamic behavior in plasmonic nanospheres.

Importantly, our calculations performed for aluminum ($\omega_{\rm p} = 14.94,\mathrm{eV}$) in Figures~\ref{fig:EELSLDOSExt}(b) and~\ref{fig:ExtEELSLDOS_1Dplot}(b), using measured data from Ref.~\citenum{Rakic:1995}, confirm the feasibility of measuring multipole resonances beyond the $l=\infty$ limit for realistic (\textit{i.e.}, non-Drude) metals: at least four orders of surface plasmons besides both dipole and quadrupole bulk plasmons are discernible.
The nanosphere radius considered in Figures~\ref{fig:EELSLDOSExt} and~\ref{fig:ExtEELSLDOS_1Dplot} is, however, relatively small at $R=1.5\,\mathrm{nm}$. While consideration of such small nanospheres eases interpretation and labeling, it also approaches the emergence of the realm of cluster physics. Nevertheless, similar spectral features persist for larger spheres, upholding the pertinence of the analysis. Supporting calculations for $R=3\,\mathrm{nm}$ nanospheres are presented in the SI.

We emphasize that one should not view the results in Figures~\ref{fig:EELSLDOSExt}(b) and~\ref{fig:ExtEELSLDOS_1Dplot}(b) as being fully representative of experiments: the semi-classical plasma-in-a-box hydrodynamic model necessarily cannot contain all relevant physics. In particular, it is known that the nonlocal blueshift of the \emph{dipolar} SPP for aluminum spheres in vacuum will be more than fully compensated by a redshift due to electronic spill-out.\cite{Mandal:2013} 

However, for higher-order multipoles we expect that the centroid of the induced charge will be pushed inwards at larger multipole orders, and that nonlocality will come to dominate the effects of spill-out. These considerations are supported by calculations in Ref.~\citenum{Liebsch:1993a} on planar simple-metal surfaces, which show that the induced charge recedes to the interior of the metal at large momentum transfers, equivalent to high multipole order. This indicates that spill-out does not undo our prediction that higher-order SPP resonances will be well-separated due to nonlocal response, and thus suggests a novel direction for identification of hydrodynamic behavior in nanospheres. The key features of our theoretical near-field spectra for aluminum are encouraging in this respect. Accordingly, experimental investigation and further theoretical study with more microscopic models is highly desirable.

Additionally, we note that electronic spill-out is not a property of the metal nanoparticle alone but also of its surrounding dielectric, in a similar way that the atomic spontaneous-emission rate is not a property of the atom alone but also of its electromagnetic environment. This gives additional experimental freedom: by embedding metal spheres into a solid matrix, electronic spill-out can be controlled and the associated redshift suppressed.\cite{vomFelde:1988} A high-index dielectric surrounding can significantly reduce the electronic spill-out, even in simple metals. Thus with high-index background dielectrics, our plasma-in-a-box model is expected to be more accurate. The key effects of a non-unity background dielectric function on the SPP and bulk-plasmon resonances of  Figures~\ref{fig:EELSLDOSExt} and~\ref{fig:ExtEELSLDOS_1Dplot} can be readily discerned from Eqs.~(\ref{eq:localresonances})\,--\,(\ref{eq:bulkplasmonapproximation}).

As further promising experiments, we propose to use the same materials as in Ref.~\citenum{vomFelde:1988}, namely potassium (or Na or Rb) nanospheres in an MgO matrix, but now for doing EELS on an individual nanosphere, so that inhomogeneous broadening would no longer obscure individual multipolar peaks.
Similarly, rather than utilizing a continuous embedding matrix, it may be feasible to suppress the electronic spill-out just by coating the nanospheres with a suitable dielectric, thereby also providing protection from oxidization.

The higher-order LSPs that we propose to observe were not seen in the recent EELS measurements on silver nanospheres of Refs.~\citenum{Scholl:2012} and \citenum{Raza:2013_Nanophotonics}. This agrees with calculations performed by us for silver, which are detailed in the SI: due to strong interband effects, higher-order multipole LSP resonances are obscured even in individual Ag nanospheres.

Above the plasma frequency, two hydrodynamic peaks can be seen in the (identical) extinction spectra of Figures~\ref{fig:EELSLDOSExt}(a) and ~\ref{fig:ExtEELSLDOS_1Dplot}(a). They clearly have no analogue in the LRA, and correspond to the first two dipolar confined bulk plasmon resonances, with labels $[l,n]=[1,1]$ and $[1,2]$, that we also identified in the hydrodynamic Mie--Lorenz coefficients in Figure~\ref{fig:MieCoefs_BulkFocus_Insets}. They have first been predicted by Ruppin to exist in the extinction spectrum.\cite{Ruppin:1973a}
Interestingly, in the EELS and LDOS spectra of Figures~\ref{fig:EELSLDOSExt} and \ref{fig:ExtEELSLDOS_1Dplot}, we see more resonances above the plasma frequency than the two dipolar bulk plasmons of the extinction spectrum. According to our investigations of the Mie--Lorenz coefficients in Figures~\ref{fig:MieCoefs_BulkFocus_Insets} and~\ref{fig:pseudodispersion}, these additional resonances in principle could be either high-$l$ LSP resonances or quadrupolar and higher-order bulk plasmon resonances. They all turn out to be bulk plasmons, and are therefore labeled accordingly; the high-$l$ LSP resonances are much weaker and absent in the spectrum.

Better than observing shifts in LSP peaks, observing the confined bulk plasmon peaks would constitute a unique identification of hydrodynamic pressure waves in nanospheres. However, since we find them to be three orders of magnitude weaker than the dipolar LSP resonance, actually the same order of magnitude weaker as found in recent density-functional calculations,\cite{Townsend:2011a} they are difficult to measure in nanospheres. To our knowledge they have not yet been observed (unlike their counterparts in thin films), so to date bulk plasmons are ``non-smoking guns'' of hydrodynamic pressure waves in nanospheres.

Overall, Figures~\ref{fig:EELSLDOSExt} and~\ref{fig:ExtEELSLDOS_1Dplot} illustrate the importance of the dimensionality of the excitation source. As is well known, the plane wave used in extinction measurements only excites dipole resonances in deeply subwavelength spheres. As to the EELS spectra, the one-dimensional source of a travelling electron excites a cylinder-like field, which for short probe-to-surface separations is sufficiently inhomogeneous to excite higher-order ($l>1$) plasmons as well.
Lastly, the LDOS spectra illustrate the largest sensitivity to the multipole modes, with all LSPs discernible and significant response from several bulk plasmon orders. The spherical-like field of the zero-dimensional dipole induces locally a more inhomogeneous excitation field than the traveling electron, thus accounting for the increased multipole-sensitivity in LDOS compared to EELS.
At large probe-surface separations shown in Figure~\ref{fig:EELSLDOSExt}, the exciting fields in both EELS and LDOS are almost homogeneous near the sphere, and the response due to higher-order multipoles is diminished. As a consequence, for large probe-to-surface separations the spectral response in extinction, EELS, and LDOS is qualitatively the same. See SI for analytical considerations of this latter point, regarding the asymptotics of the LDOS and EELS spectra.

\begin{figure}
\centering
\includegraphics[]{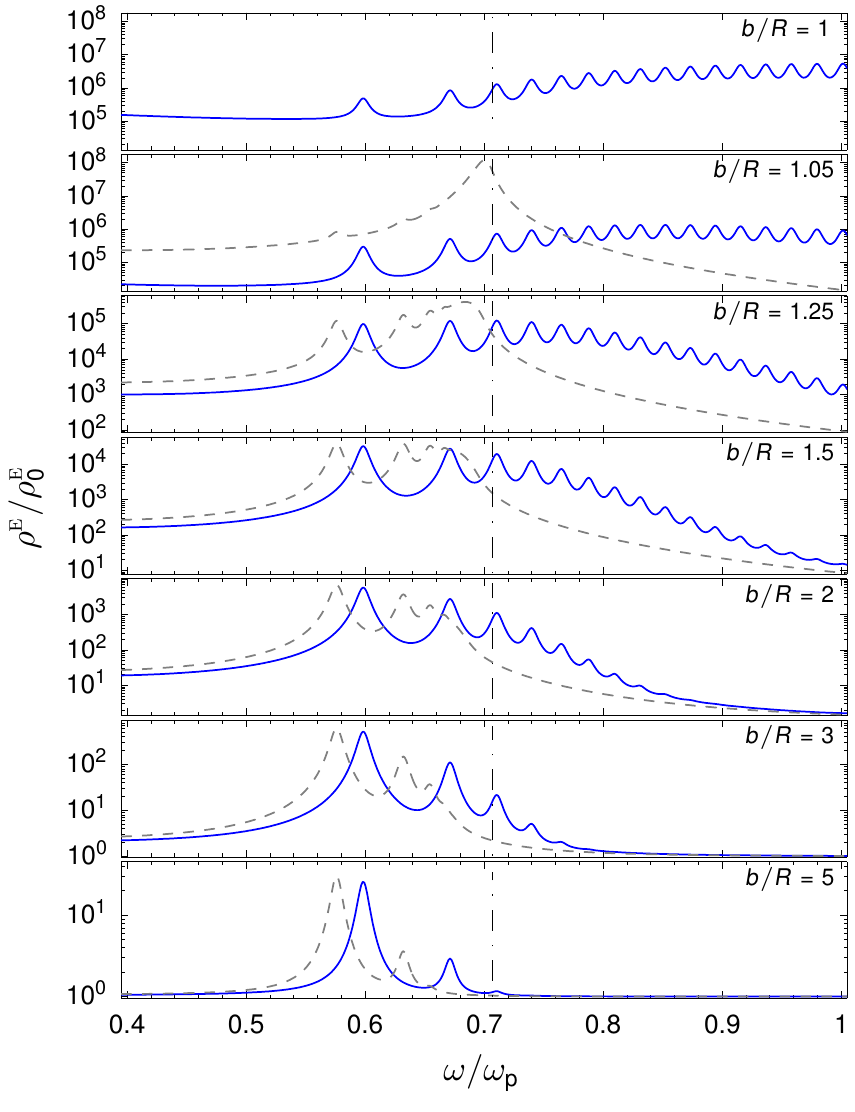}
\caption{Normalized LDOS for different probe-to-surface separations in hydrodynamic and LRA treatments, in full blue and dashed gray, respectively, for a Drude-metal with material parameters as in Figure~\ref{fig:EELSLDOSExt}(a), for a $R= 2.5\, \mathrm{nm}$ sphere.}\label{fig:LDOS_bvar}
\end{figure}

\emph{Distance-dependence of LDOS.}\quad{}
In the preceding sections we established that the response of high-order plasmons is significantly enhanced with probes of low-dimensionality when examined in the near-field, where the observability of multipolar LSPs is enhanced by hydrodynamics itself. Let us therefore finally focus solely on the LDOS spectra, where the response of these high-order multipoles is most pronounced.
In Figure~\ref{fig:LDOS_bvar}
we display the variation of the LDOS spectra as a function of the probe-to-surface separation, varying from $b/R=1$ (\textit{i.e.}, source on surface) to $b/R=5$ ($10\,\mathrm{nm}$ separation). For the panels with $b/R < 2$, contributions from high-order multipoles are increasingly important, as the excitation of multiple LSP orders contribute to the spectrum\cite{Moroz:2010}. Consequently, in the LRA the largest LDOS occurs at $\omega_{\rm p}/\sqrt{2}$, the limiting frequency of the high-order LSPs, coinciding with the pile-up of LRA multipoles. By contrast, the hydrodynamically blueshifted LSPs do not have a finite limiting frequency or an associated similar pile-up of modes, but instead exhibit distinguishable peaks associated with excitation of different multipoles. The qualitative discrepancy between local and nonlocal spectra is even substantial. For larger spheres, the multipole peaks merge and instead give rise to a broadband enhancement above $\omega_{\mathrm{p}}/\sqrt{2}$, even extending beyond the plasma frequency, see SI for supporting calculations on an $R=10\,\mathrm{nm}$ sphere. This suggests an hitherto largely unexplored regime of studying nonlocal response in comparatively large nanostructures but at short surface-to-probe separations.

As is well known, in the extreme limit $b=R$ the LRA LDOS diverges (hence not shown) due to the $1/(b-R)^3$ scaling of the nonradiative decay rate. For the $b/R=1$ panel of Figure~\ref{fig:LDOS_bvar} we obtain convergent results for the hydrodynamic response, and associated finite LDOS spectra. The convergence, however, hinges upon the choice of a simple Drude-metal with real-valued $\varepsilon_{\infty}$, as discussed in Ref.~\citenum{Datsyuk:2011}. As such, hydrodynamic response does not fully regularize the divergence of the LDOS for real metals with dissipative bound response. Complete regularization in real metals would likely necessitate an appropriate nonlocal treatment of not only the free response, but also the bound response. In addition, for these very close proximities between source and nanosphere, the effect of high-order moments - beyond the dipole -- of the source itself, due to the finite size of the source, would certainly modify the decay rates as well.\cite{Andersen:2011a} For emitters at the larger separations, \textit{e.g.}, in the panel with $b/R = 5$, the dipole mode of the nanosphere is again the primary feature, but with the quadrupolar LSP still imposing a significant spectral feature.

\section{Conclusions}
In this paper we have aimed to identify indisputable signatures of hydrodynamic response in plasmonic nanospheres.
The corresponding evidence for layered systems is the observation, found both with light\cite{Lindau:1971,Anderegg:1971} and with electrons,\cite{Oezer:2011} of confined bulk plasmons in thin films. Employing the hydrodynamic Drude model we predict the existence of confined bulk plasmons also in nanospheres. An important question then, is whether such excitations would be observable. A series of confined bulk plasmons of dipolar character has been predicted before to show up in extinction spectra.\cite{Ruppin:1973a} Here we additionally found that besides the dipole series, also series of quadrupolar and higher-order bulk plasmons emerge in near-field EELS and LDOS spectra. However, we find the strength of these bulk plasmon resonances in nanospheres to be about three orders of magnitude weaker than the dominant LSP peak. Their experimental observation in nanospheres, for example with EELS or LDOS, remains an open challenge. Another promising technique is core-level photoemission.\cite{Oezer:2011}

Of a more immediate, accessible nature experimentally, is our prediction that in the near-field EELS and LDOS spectra also quadrupolar and higher-order LSPs appear, besides the well-known dominant dipolar LSPs. In itself it is no surprise that higher-order LSPs show up in near-field spectra, since already the LRA predicts them\cite{Moroz:2010}. The salient point here is that LRA LSPs exhibit the surface plasmon $\omega_{\mathrm{p}}/\sqrt{2}$ of a planar interface as a limiting upper frequency, while we predict hydrodynamic LSPs to be observable also above $\omega_{\mathrm{p}}/\sqrt{2}$. This follows from our prediction that higher-$l$ plasmons exhibit a larger nonlocal blueshift. Indeed, we found that high-$l$ LSPs in principle can occur above the plasma frequency in few-nanometer spheres, with their mode profiles still well bound to the surface. An upper limiting frequency for LSPs actually does not exist in the hydrodynamic model.

Not all multipolar LSPs will be observable, though. For silver, we predict all LSPs besides the dipolar one to be suppressed due to interband effects. On the other hand,  we predict that for aluminum nanospheres several higher-order LSPs should be observable in near-field EELS and LDOS spectra. In ensembles of alkali metal (Na, K, Rb) nanospheres in an MgO matrix, resonances above the LRA limit $\omega_{\mathrm{p}}/\sqrt{2}$ have actually already been observed, but individual resonance peaks could not be resolved due to ensemble averaging.\cite{vomFelde:1988} We propose to do these measurements on individual alkali metal nanospheres, something that has already been achieved with silver nanospheres.\cite{Scholl:2012,Raza:2013_Nanophotonics}

Would such measurements constitute the unequivocal evidence, the ``smoking gun'', of hydrodynamic nonlocal response in nanospheres that we set out to identify? We can only suggest `perhaps' at this stage, because alternative explanations for resonances above $\omega_{\mathrm{p}}/\sqrt{2}$ do exist. In particular, vom Felde \textit{et al.} invoke quantum confinement (cluster physics) rather than hydrodynamics (nanoplasmonics) to explain their intriguing observation of resonances above the LRA limit.\cite{vomFelde:1988} It is safe to assume, however, that fitting the two distinct models to a measured series of LSP resonances will be more conclusive than fitting only the dominant dipolar LSP, which remains state of the art.\cite{Scholl:2012,Raza:2013_Nanophotonics,CarminaMonreal:2013a} We therefore suggest to measure near-field EELS and LDOS spectra of nanospheres of aluminum and alkali metals embedded in a solid dielectric environment.

The plasmonic resonances emerge with strikingly different weights in the three types of spectra that we calculated, so that for example the state-of-the-art comparison of EELS experiments with theoretical absorption cross sections\cite{Scholl:2012} or extinction cross sections\cite{Raza:2013_Nanophotonics} can be of limited use. The comparison happened to be useful for silver nanospheres,\cite{Scholl:2012,Raza:2013_Nanophotonics} where interband effects suppress the beyond-dipole LSP resonances that otherwise would show up in near-field EELS and LDOS experiments.

Even for the relatively simple hydrodynamic theory that we used here, the near-field spectra of nanospheres become rather elaborate and rich -- but they can be understood rigorously. We therefore expect that our results could also assist in the interpretation of near-field spectra calculated with more microscopic calculations, with some features attributable to hydrodynamic nonlocal response.

%
%
\section{Methods}
\begin{small}
\paragraph{Hydrodynamics and multipole basis.}
By eliminating the current density in Eqs.~\eqref{eqs:governing}, the hydrodynamic equations can be recast solely in terms of the electric field:
\begin{subequations}\label{eqs:hydroJgrp}
\begin{align}
\left(\nabla^2 + k_{\textsc{m}}^2 \right) \nabla \times \boldsymbol{\mathrm{E}}(\boldsymbol{\mathrm{r}},\omega) &= 0, \label{eq:hydroET} \\
\left(\nabla^2 + k_{\textsc{nl}}^2 \right) \nabla \cdot \boldsymbol{\mathrm{E}}(\boldsymbol{\mathrm{r}},\omega) &= 0, \label{eq:hydroEL}
\end{align}
\end{subequations}
where $k_\textsc{m}^2 = k_0^2 \varepsilon_{\textsc{m}}$ and $k_{\textsc{nl}}^2 = (\omega_{\mathrm{p}}/\beta_{\textsc{f}})^2 \varepsilon_{\textsc{m}}/[\varepsilon_{\infty}(\varepsilon_{\infty}-\varepsilon_{\textsc{m}})]$ 
denote the transverse and longitudinal wavenumbers in the metal, respectively. The transverse response of the metal is governed by $\varepsilon_{\textsc{m}}(\omega) = \varepsilon_\infty(\omega) - \sigma(\omega)/i\varepsilon_0\omega$.

The vector wave functions, $\mathbf{M}_{\nu}(\boldsymbol{\mathrm{r}})$, $\mathbf{N}_{\nu}(\boldsymbol{\mathrm{r}})$, and $\mathbf{L}_{\nu}(\boldsymbol{\mathrm{r}})$, are defined in terms of a pilot vector $\mathbf{c}$, and a generating scalar function $\psi_{\nu}(\boldsymbol{\mathrm{r}})$, satisfying the Helmholtz equation $\nabla^2\psi_{\nu}(\boldsymbol{\mathrm{r}}) + k^2\psi_{\nu}(\boldsymbol{\mathrm{r}}) = 0$. In spherically symmetric structures it is natural to express the generating functions in spherical coordinates $\boldsymbol{\mathrm{r}} = (r,\theta,\phi)$ and to choose the pilot vector as the (non-constant) outward radial vector $\mathbf{c} = \boldsymbol{\mathrm{r}}$. In this case, the degeneracy label $\nu$ separates into the angular momentum quantum numbers $l$ and $m$, and the vector wave functions read as
\begin{subequations}\label{eqs:SphericalVectorWaves}
\begin{align}
\mathbf{M}_{lm}(\boldsymbol{\mathrm{r}}) &= \nabla \times \mathbf{r}\psi_{lm}(\boldsymbol{\mathrm{r}}),\\
\mathbf{N}_{lm}(\boldsymbol{\mathrm{r}}) &= \frac{1}{k}\nabla \times \nabla \times \mathbf{r}\psi_{lm}(\boldsymbol{\mathrm{r}}),\\
\mathbf{L}_{lm}(\boldsymbol{\mathrm{r}}) &= \frac{1}{k}\nabla \psi_{lm}(\boldsymbol{\mathrm{r}}),
\end{align}
\end{subequations}
with $\psi_{lm}(r,\theta,\phi) = z_l(kr)P_l^m(\cos\theta)\mathrm{e}^{im\phi}$, where $z_l$ denotes spherical Bessel or Hankel functions of the first kind, $j_l$ or $h_l^{\scriptscriptstyle (1)}$, for in- and outgoing waves, respectively. Finally, $P_l^m$ denote the associated Legendre polynomials. In addition, by requirements of continuity along $\phi$ and boundedness at the polar extremes, the angular momentum quantum numbers are restricted to integer values in the ranges $l\in [1,\infty[$ and $m\in[-l,l]$. This particular basis is usually referred to as the multipole basis.

The $k$-dependence of the vector wave functions used in the field expansions varies inside and outside the sphere. By insertion of the external field into the vector Helmholtz equation, $\nabla^2\boldsymbol{\mathrm{E}} + k_{\textsc{d}}^2\boldsymbol{\mathrm{E}} = 0$, which is valid outside the sphere, it is clear that the appropriate choice of wavenumber is $k_{\textsc{d}} = \sqrt{ \varepsilon_{\textsc{d}} }k_0$ outside the sphere. Similarly, by insertion of the internal field into Eqs.~\eqref{eqs:hydroJgrp}, it is clear that the solenoidal vector waves $\mathbf{M}_{lm}^{\mathrm{tr}}$ and $\mathbf{N}_{lm}^{\mathrm{tr}}$ inside the sphere are associated with the transverse wavenumber $k_{\textsc{m}}$, while the irrotational vector wave $\mathbf{L}_{lm}^{\mathrm{tr}}$ is associated with the longitudinal wavenumber $k_{\textsc{nl}}$.

Finally, the matching of internal and external expansions is facilitated by application of BCs. The usual BCs for the electromagnetic field requires the continuity of the tangential components of the electric and magnetic field at $r=R$, \textit{i.e.}, $\boldsymbol{\mathrm{E}}^{\mathrm{ex}}_{\scriptscriptstyle \parallel}+ \boldsymbol{\mathrm{E}}^{\mathrm{sc}}_{\scriptscriptstyle \parallel} = \boldsymbol{\mathrm{E}}^{\mathrm{tr}}_{\scriptscriptstyle \parallel}$ and $\mathbf{H}^{\mathrm{ex}}_{\scriptscriptstyle \parallel}+ \mathbf{H}^{\mathrm{sc}}_{\scriptscriptstyle \parallel} = \mathbf{H}^{\mathrm{tr}}_{\scriptscriptstyle \parallel}$. Furthermore, an additional BC is required to account for the presence of the longitudinal waves inside the metal, which, in the case of an abrupt dielectric boundary, is unambiguously chosen as the continuity of the normal component of the induced current, equivalent to the continuity of the normal component of the bound-charge depolarization at $r=R$, corresponding to $\varepsilon_\textsc{d}\boldsymbol{\mathrm{E}}^{\mathrm{ex}}_{\scriptscriptstyle \perp}+ \varepsilon_\textsc{d}\boldsymbol{\mathrm{E}}^{\mathrm{sc}}_{\scriptscriptstyle \perp} = \varepsilon_{\infty}\boldsymbol{\mathrm{E}}^{\mathrm{tr}}_{\scriptscriptstyle \perp}$ \cite{Boardman:1982a,Wei_PRB2012}.
\end{small}
%
%

\begin{acknowledgement}
The Center for Nanostructured Graphene is sponsored by the Danish National Research Foundation, Project DNRF58. This work was also supported by the Danish Council for Independent Research - Natural Sciences, Project 1323-00087.
\end{acknowledgement}

\begin{suppinfo}
Additional information regarding computation of extinction cross-section, EELS probability, and LDOS in the multipole basis along with asymptotics of EELS and LDOS in near-extinction limits. Also provides derivation of the quasistatic multipolar polarizability, associated LSP resonance conditions, Mie--Lorenz transmission coefficients, approximate longitudinal resonance conditions for bulk plasmons, quasistatic LDOS expressions and associated calculations for $R=10\,\mathrm{nm}$ Drude-metal sphere, and supplementary calculations for silver and larger nanospheres.
\end{suppinfo}

\mciteErrorOnUnknownfalse 


\providecommand*\mcitethebibliography{\thebibliography}
\csname @ifundefined\endcsname{endmcitethebibliography}
  {\let\endmcitethebibliography\endthebibliography}{}

\end{document}